\documentstyle[aps]{revtex}
\input{psfig}

\begin{document}

\textheight 8.8in
\textwidth 6.5in
\topmargin -.25in
\oddsidemargin -.25in
\evensidemargin 0in
\baselineskip 14pt
\def\hm{\ \rm {\it h}^{-1} Mpc}

\title{Perturbations in Quintessential Inflation}

\author{Carlo Baccigalupi$^{1}$ and Francesca Perrotta$^{2}$}
\address{
$^{1}$ INFN and Dipartimento di Fisica, Universit\`a di Ferrara, 
Via del Paradiso 12, 44100 Ferrara, Italy;\\
$^{2}$ SISSA/ISAS, Via Beirut 4, 34014 Trieste, Italy.
}

\baselineskip 10pt
\maketitle
%\today
\begin{abstract}
Recently Peebles and Vilenkin proposed and quantitatively 
analyzed the fascinating idea that a substantial fraction of the 
present cosmic energy density could reside in the vacuum potential energy 
of the scalar field responsible for inflation (quintessential inflation). 
Here we compute the signature of this model in the cosmic microwave 
background polarization and temperature anisotropies and in the large 
scale structure. 
\end{abstract}

\section{Introduction}

In the last decade the connection between cosmology and 
particle physics has become more and more interesting. 
We have several sectors in which a comparison between 
the theoretical high energy physics processes 
of the very early universe and their observable traces is possible. 
These research frontiers include the exploration of the 
anisotropies in the cosmic microwave background (CMB) 
and of the large scale structure in the present matter distribution
(LSS); wide and high resolution experiments designed to 
gain observationally insight into these topics are currently 
in preparation \cite{CMB,LSS}. 

Let us approach the subject of this work. Very recently 
the emissions from deep type Ia supernova 
have become observable with unprecedented high resolution 
\cite{IA}. The surprising news arising from 
these observations suggest that we are currently 
living in a universe that is {\it accelerating} its expansion. 
As it is well known, this could be the observable 
effect of a vacuum energy density comparable with 
the critical one. A possible explanation could be 
the existence of a cosmological constant, much smaller 
than the characteristic energy scales of quantum gravity 
effects but non-zero. This {\it ad\ hoc} possibility is unlikely for
theorists, tending to believe that some unknown process set the 
cosmological constant to zero in the very early universe \cite{CC}. 

Recently the idea that the vacuum energy density could 
be mimicked by a dynamical scalar field $\phi$, named 
quintessence, has been considered 
with more and more interest since it provide several nice 
features in the CMB power spectrum and LSS. More precisely, 
this occurs in the two most popular candidates proposed so far 
for this field, a cosine potential for 
the Pseudo Nambu Goldstone Boson \cite{CDF} and 
an exponential potential \cite{FJ}. 

However, since most of the inflationary phenomenology is based 
on the dynamics of a scalar field , the inflaton, it is tempting 
to relate quintessence to inflaton. A first proposal 
from the scientific community in this sense was made very 
recently \cite{PV}. The authors suggested that quintessence and 
inflaton are the same field seen at different times. 
They proposed a detailed model, providing an appropriate set of 
values for the physical constants in order to realize this 
fascinating possibility. The key feature is the occurring of 
a kinetic energy dominated phase that connects inflation and 
radiation era. Although the model has to be further 
investigated, especially for what concerns the 
spectrum of gravitational waves and the possible 
production of topological defects, it is interesting to look at 
the perturbations dynamics induced by its basic features. 
A quintessence model with similar characteristics is being 
considered by other authors \cite{ZWS}, with emphasis on 
arguments regarding the viability of general quintessence 
models. In this work we concentrate on the power spectra of 
CMB polarization and temperature anisotropies as well as 
on the LSS perturbations produced in this scenario. 
In section II we recall the background dynamics in 
quintessential inflation and in section III we describe 
its perturbations; finally, in section IV we 
numerically compute and discuss the signature of this scenario 
on the CMB polarization and temperature power spectra and 
on the LSS. 

\section{quintessential inflation}

In this section we briefly review the basic features of 
quintessential inflation; the reader is advised to 
look at the original work in \cite{PV} for a complete exposition 
and references. 

The model involves a minimally coupled scalar field 
with potential 
$$
V(\phi )=\lambda\cdot (M^{4}+\phi^{4})\ \ {\rm for}\ \ \phi < 0\ ,
$$
\begin{equation}
V(\phi )={\lambda M^{8}\over (M^{4}+\phi^{4})}
\ \ {\rm for}\ \ \phi\ge 0\ .
\label{pvp}
\end{equation}
Inflation occurs for $\phi\ll -M$; radiation and 
matter dominated eras for $\phi\gg M$. In order 
to make particles and perturbations production 
just as in chaotic inflationary models \cite{MFB}, 
and to have about $70\%$ of the critical energy today in 
quintessence, the following physical constants, 
in $\hbar =1,c=1$ units, have been chosen: 
\begin{equation}
\lambda= 10^{-14}\ \ ,\ \ M=8\cdot 10^{5}\ {\rm GeV}\ .
\label{pc}
\end{equation}
The cosmic trajectory is assumed to begin at $\phi\ll -M_{PL}$, 
with an era of chaotic inflation. At $\phi\simeq -M_{PL}$ 
a kinetic energy dominated era begins; in this epoch 
the (kinetic) quintessence energy density decreases 
very rapidly, $\rho_{\phi}\sim a^{-6}$ where $a$ is 
the scale factor. As in the ordinary scenario, particles 
are produced in the curved space-time from an initial 
quantum vacuum state \cite{MFB}. In order to have a  workable model, 
it is necessary that the kinetic era ends when the total field 
energy is negligible with respect to the radiation one; 
if this is the case, the radiation era begins and $\phi$ starts 
its slow rolling toward the present state. 
It is interesting to note that this model gives a 
reheating temperature curiously comparable with 
the supposed electroweak symmetry breaking scale, 
$T_{rh}\simeq 10^{3}N_{\psi}^{3/4}$GeV, where 
$N_{\psi}$ is the number of scalar fields involved 
in the process. 

This is the general phenomenology imposed by the constants 
(\ref{pc}). At the present time, in the matter dominated era, 
the inflaton is totally equivalent to a quintessence field, 
rolling on the potential $V(\phi )\simeq \lambda M^{8}/\phi^{4}$ 
with a simple time evolution, $\phi\simeq 2^{1/3}
\lambda^{1/6}M^{4/3}H_{0}^{-1/3}/\sqrt{1+z}$. 

\section{Perturbations}

Perturbations in models with a dynamical scalar field 
together with the other ordinary matter and 
radiation particles require a generalization \cite{PB} of 
earlier works \cite{MB}; a complete treatment of this subject 
can be found in the cited works and here we report only the 
relevant issues for the present problem. 

Even if the reheating temperature in the present scenario is much 
smaller than in chaotic inflation, radiation dominates well before 
nucleosynthesis. In the most simple view, the cosmic fluid 
can be thought composed by photons ($\gamma$), 
baryons ($b$), cold dark matter ($cdm$) and 
three families of massless neutrinos ($\nu$). 
As we briefly exposed in the previous section, Gaussian 
perturbations arise adiabatically from the inflaton dynamics 
at the end of inflation \cite{PV}. They involve matter and 
radiation as well as fluctuations $\delta\phi$ of the scalar field 
around its background value $\phi$. 

The initial conditions for the perturbations are posed at early 
conformal time $\tau =\int_{0}^{t}dt/a(t)$ when essentially all 
the perturbation wavenumber $k$ interesting for structure 
formation are well outside the effective horizon, $k\tau\ll 1$. 
Adiabatic conditions are posed initially by requiring that no 
gauge invariant entropy perturbation difference exists between 
any pair of components \cite{PB}; 
in the conformal Newtonian gauge, the leading order 
early time behaviour for the scalar quantities evolving 
from adiabatic initial conditions is 
$$
\delta_{\gamma}={4\over 3}\delta_{b}=
{4\over 3}\delta_{cdm}=\delta_{\nu}\propto constant\ ,
$$
$$
v_{\gamma}=v_{b}=v_{cdm}=v_{\nu}\propto k^{2}\tau\ \ ,
\ \ \sigma_{\nu}\propto k^{2}\tau^{2}\ ,
$$
\begin{equation}
\delta\phi\propto \left({d\phi\over dt}\right)_{t=0}
\cdot\tau^{2}\ ,
\label{ic}
\end{equation}
where $\delta,v,\sigma$ means density, velocity and shear 
perturbations respectively. Note that $\delta\phi$ is initially 
linked to the kinetic field energy. 
A multipole expansion accounts for 
temperature as well as polarization perturbations 
of the Planckian black body spectrum arising mainly from 
Thomson scattering at the energies relevant in the present problem; 
neutrinos also are treated similarly without the Thomson 
scattering terms \cite{MB}. 

From this initial regime, perturbations evolve 
according the linearized Einstein and Boltzmann 
equations in a flat Friedmann Robertson Walker 
background; the latter involves quintessence, scale factor and 
unperturbed density of all the fluid species, 
being driven by the Klein Gordon and unperturbed Einstein 
equations respectively (see \cite{PB} and references therein). 

\section{Results and discussion} 

We require that at the present about $70\%$ of the 
critical energy density resides in quintessence. 
This energy comes essentially from the potential 
component, since $\phi$ is rolling very slowly making 
the kinetic energy negligible, as we show below; the 
baryon abundance respects the nucleosynthesis constraint: 
\begin{equation}
\Omega_{\phi}=.7\ \ ,\ \ \Omega_{b}=.05
\ \ ,\ \ \Omega_{cdm}=1-\Omega_{\phi}-\Omega_{b}\ .
\label{o}
\end{equation}
Also we adopt $H_{0}=70$ km/s/Mpc consistently with some 
present measurements \cite{HUBBLE} and assume an 
initial power spectrum exactly scale-invariant. 

The request that the present amount of quintessence energy 
is $\Omega_{\phi}$ does not fix completely its dynamics; since 
it obeys the Klein Gordon equation we need to specify 
its time derivative. At the present this is very low 
since it has been redshifted away during the expansion occurred 
in the radiation and matter eras (assuming that it is not too large 
with respect to the potential energy at the beginning of the 
radiation dominated era). However, particularly in this 
scenario where the kinetic energy plays a fundamental 
role during the cosmic evolution, it is important to 
take into account the field time derivative. 
This is realized in the following way: after fixing 
$\Omega_{\phi}$, the code asks for the initial 
kinetic to potential energy ratio; the sign of the time 
derivative is then chosen toward the direction of lower 
potential of course, that is $d\phi /dt >0$ in this case. 
In all the cases analyzed $\phi$ has initially an 
equal amount of kinetic and potential energy. 

We examine the imprint on the main observational topics 
for this scenario. We show our results regarding  
CMB polarization, temperature and linear matter power spectrum in 
figures 1,2,3 respectively. In these figures, solid curves
having  increasing amplitudes describe quintessence models having 
increasing values of $\Omega_{\phi}$. 

Polarization anisotropies in the CMB has not yet measured; the 
existing upper limits are at the level of the measured temperature 
anisotropies, see \cite{W} and references therein for further 
details. On the other hand it is expected to arise naturally as 
the result of the anisotropic nature of the Thomson scattering; 
thus polarization anisotropies arises mainly from the CMB acoustic 
oscillations occurring on sub-horizon scale at decoupling, 
that is a degree in the sky or less. Figure 1 shows the 
power spectra of the polarization anisotropies in quintessential 
inflation (solid line). The thin dashed line represents an 
ordinary Cold Dark Matter model (CDM) in which the energy 
density associated with quintessence has been replaced with 
dark matter. The amplitude of the peaks 
increases and there is a global shift toward higher multipole 
indexes, or smaller scales. The first effect is due to the 
lack of matter at decoupling with respect to the CDM model: 
the universe at decoupling is mildly radiation dominated 
and this enhances the radiation perturbation amplitude 
(see \cite{PB} and references therein). The second is a projection 
 effect  resulting  from  a pure geometric feature \cite{HSS};
 the comoving distance 
of the last scattering surface in quintessence 
models is larger than in CDM, thus shifting the angular 
scales corresponding to the coherent acoustic oscillations 
toward smaller angular scales. 

The same features occur in the COBE normalized CMB temperature 
power spectra shown in figure 2. In this case a third effect 
arises from the integrated Sachs-Wolfe effect due to the time 
evolution of matter perturbations along the photons path. 
Again this is due to the lack of matter in quintessence model 
with respect to the CDM. Also this is effective mainly on 
super-horizon scales, low multipoles in the figure, that prevent the 
cancelation from the oscillatory sub-horizon dynamics. 

We used linear perturbation theory to calculate the
 matter  power spectra $ \ P(k) \ $ plotted in figure 3.
 They are defined by 
$< \delta ( {\bf k})  {\delta}^{\ast} ( {\bf k'})> =
 4 \pi P(k) {\delta}_D ( {\bf k-k'})$, where ${\delta}_D$ is the 
Dirac delta function and $ \delta ( {\bf k})$ is the Fourier transform 
of the spatial matter density fluctuation field. 

The epoch of matter-radiation equality in quintessence scenarios
is obviously closer to the present 
 than in CDM models, because of the lack of matter. 
The effective horizon scale at equivalence corresponds roughly 
to the location of the turnover in the  spectra; 
this causes the shift 
toward larger scales, or small wavenumbers, and therefore 
subtracts power to the small scale structure as indicated 
by the current data from galaxy surveys \cite{RS}. \\
The dispersion of the density field is quantified by values of
$\sigma_8 = .59$ for $\Omega_{\phi}=.8$,
$\sigma_8 = .87$ for $\Omega_{\phi}=.7$,
$\sigma_8 = 1.08$ for $\Omega_{\phi}=.6$, 
which are to be compared with standard CDM ($\Omega_{matter}=1$) model
prediction of $\sigma_8 = 1.63$.

Before concluding, it is necessary to point out here that most of 
the dynamics of background and perturbations regards the 
$\phi\ge 0$ side of the potential (\ref{pvp}). In other words, 
in the present analysis the distinctive features on CMB and 
LSS come from the form of the potential in the 
radiation and matter dominated eras and from the assumption 
of initial adiabaticity. If the future observed spectra should be 
in agreement with the ones computed here, we shall be able to state 
that the observed cosmology is consistent with 
quintessential inflation, without confirming it 
definitely. Surely this model has to be further investigated, 
and hopefully its predictions will be enriched by a more 
deep understanding of its phenomenology at the transition between 
inflation and radiation era; ultimately this scenario will be 
further constrained by experimental enterprises of the next 
generation, beginning from the primordial gravitational waves 
spectrum. 

\acknowledgements

We are grateful to Sabino Matarrese for his warm encouragement.

\begin{figure}
\psfig{figure=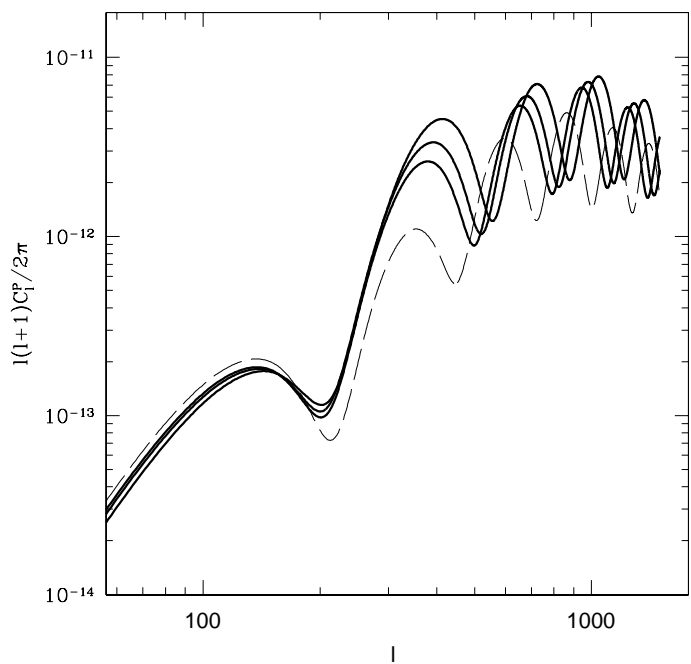,height=5in}
\caption{Polarization power spectra in quintessential 
inflation (solid lines) compared with the CDM model 
(thin dashed line). As in the following figure, 
solid curves with increasing amplitude represent 
$\Omega_{\phi}=60\%,\ 70\%,\ 80\%$ respectively.}
\end{figure} 
\begin{figure}
\psfig{figure=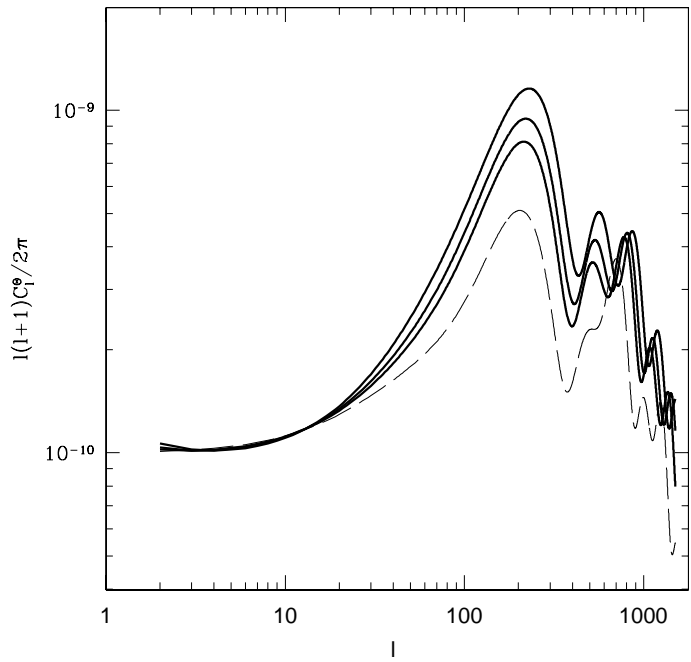,height=5in}
\caption{COBE normalized CMB power spectra in quintessential 
inflation and CDM.}
\end{figure} 
\begin{figure}
\psfig{figure=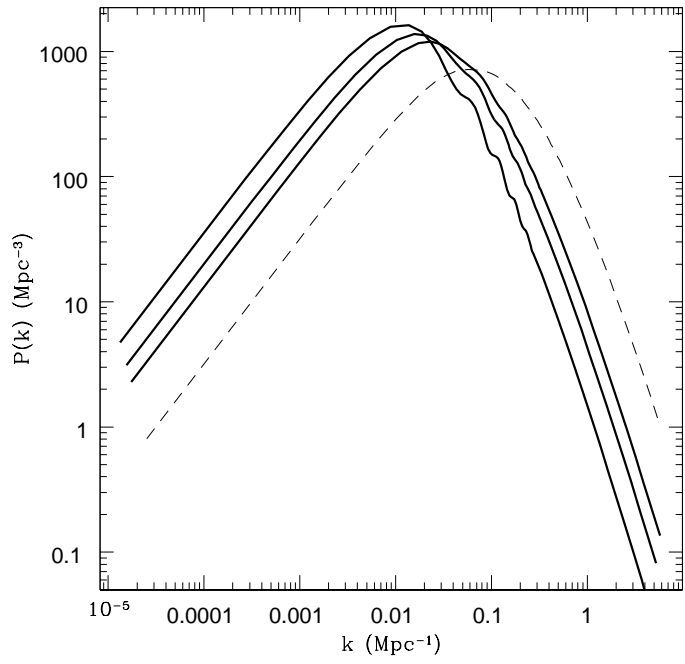,height=5in}
\caption{Large scale structure in quintessential inflation 
and CDM models.}
\end{figure} 

\end{document}